\documentclass[letterpaper, 10 pt, conference]{IEEEtran}  

\IEEEoverridecommandlockouts                              
\usepackage{graphicx}
\usepackage{graphics} 
\usepackage{epsfig} 
\usepackage{mathptmx} 
\usepackage{times} 
\usepackage{amsmath} 
\usepackage{amssymb}  
\usepackage{url}
\usepackage{xcolor}

\title{\LARGE \bf
Performance Prediction of Hub-Based Swarms
}

\author{Puneet Jain*, Chaitanya Dwivedi**, Vigynesh Bhatt*, Nick Smith*, Michael A Goodrich*
\thanks{Authors are with the Department of Computer Science,
        Brigham Young University, Provo, USA. 
        For any questions, email:
        {\tt\small puneetj@byu.edu}, {\tt\small mike@cs.byu.edu}}%
}

\begin{document}

\maketitle
\thispagestyle{empty}
\pagestyle{empty}

\begin{abstract}
A hub-based colony consists of multiple agents who share a common nest site called the hub. Agents perform tasks away from the hub like foraging for food or gathering information about future nest sites. Modeling hub-based colonies is challenging because the size of the collective state space grows rapidly as the number of agents grows. This paper presents a graph-based representation of the colony that can be combined with graph-based encoders to create low-dimensional representations of collective state that can scale to many agents for a best-of-N colony problem.  We demonstrate how the information in the low-dimensional embedding can be used with two experiments. First, we show how the information in the tensor can be used to cluster collective states by the probability of choosing the best site for a very small problem. Second, we show how structured collective \textit{trajectories} emerge when a graph encoder is used to learn the low-dimensional embedding, and these trajectories have information that can be used to predict swarm performance. 

\end{abstract}

\section{Introduction}


Biological inspiration drawn from honeybees, ants, birds, and various animal species has been
instrumental in agent-based models (ABMs) of multi-agent swarms. In ABMs, each agent independently implements its own controller, and collective behavior emerges from interactions among the agents~\cite{bonabeau1999swarm,reina2015design, reynolds1987flocks, seeley2001nest}.
ABMs capture the decentralized and individualized nature of interactions in complex systems, making them valuable for empirically studying emergent behaviors and system-level dynamics. 
This paper addresses the best-of-N problem, where agents stationed at a central hub make a distributed decision to choose the best site from a set of $N$~possibilities~\cite{valentini_best-n_2017}.


A common bottleneck for understanding large-scale bio-inspired swarms is the agents' slow decision making and the huge complexity of the system. A solution to this problem is to use differential equations~\cite{sumpter2010collective, bussemaker1997mean, nevai2010stability}, but those assume infinite agents and time. While differential equation have proven effective for generating metrics about performance of a swarm, understanding the performance of hub-based agent colonies with finite robots remains a
challenge~\cite{wang2009search, jia2013control, adams2018swarm}. 

This paper represents the collective state in an ABMs with the nodes in a graph~\cite{jaingoodrich2023}. Changes in collective state are represented as probabilistic transitions, forming a Markov chain that can be used to predict performance and other swarm properties. Unfortunately, the number of nodes, node features, and edges grows very quickly with the number of agents. This paper shows that low dimensional graph embeddings provide useful information that support computationally feasible ways of understanding swarm behavior.

The main contributions of the papers are: First, a graph-based representation is created from a relational database in which each database record encodes an individual agent's internal state. Second, the records in the relational database are ``stacked" into tensors to form a probabilistic graph of collective state behavior. Third, a graph-based encoder is constructed. Finally, the resulting node embeddings are shown to provide insight into the swarm behavior that can be applied to various swarm and problem configurations.

\section{Related Work}
Many bio-inspired swarms exhibit spatial swarming patterns such as flocking or cyclic behavior~\cite{reynolds1987flocks,couzin2002collective}. Other types of swarms are organized as hub-based colonies where all agents belong to a common nest and fan out from the nest in search of food or new suitable nest sites~\cite{adams2018swarm,reina2017model,valentini_achieving_2017}.  Swarms can be implemented as ABMs, which are employed for designing complex systems~\cite{coppola2019pagerank, parpinelli2002data, dorigo1999ant,rosenfeld2016optimal,sakellariou}. ABMs are used in social sciences to model the interactions of individuals~\cite{kimura_extracting_2010, zhang_preference-based_2011}, simulate decision making~\cite{ferguson1989solved}, and study traffic flow~\cite{mehar2013optimization}. They are also applied to understand ecosystems and biodiversity~\cite{estes1974social} and to model disease spread~\cite{omic2008virus}.


Frameworks to extend ABMs built on finite state machines to graph representations include~\cite{jaingoodrich2023,mesbahi2010graph}. Graph neural networks have also been used with multi-agent systems such as in traffic engineering~\cite{bernardez2023magnneto} and trajectory prediction~\cite{mo2021heterogeneous}. Other methods which learn on subgraphs to learn graph representation include~\cite{hamilton2017inductive, nguyen2018learning}.




\section{Graphs for the Best-of-N Problem}

This section presents our ABM formulation of the best-of-N problem and how we create graphs from the ABM.

\subsection{Best-of-N Problem}
\begin{figure}[h]
      \centering
    \includegraphics[width=0.8\linewidth]{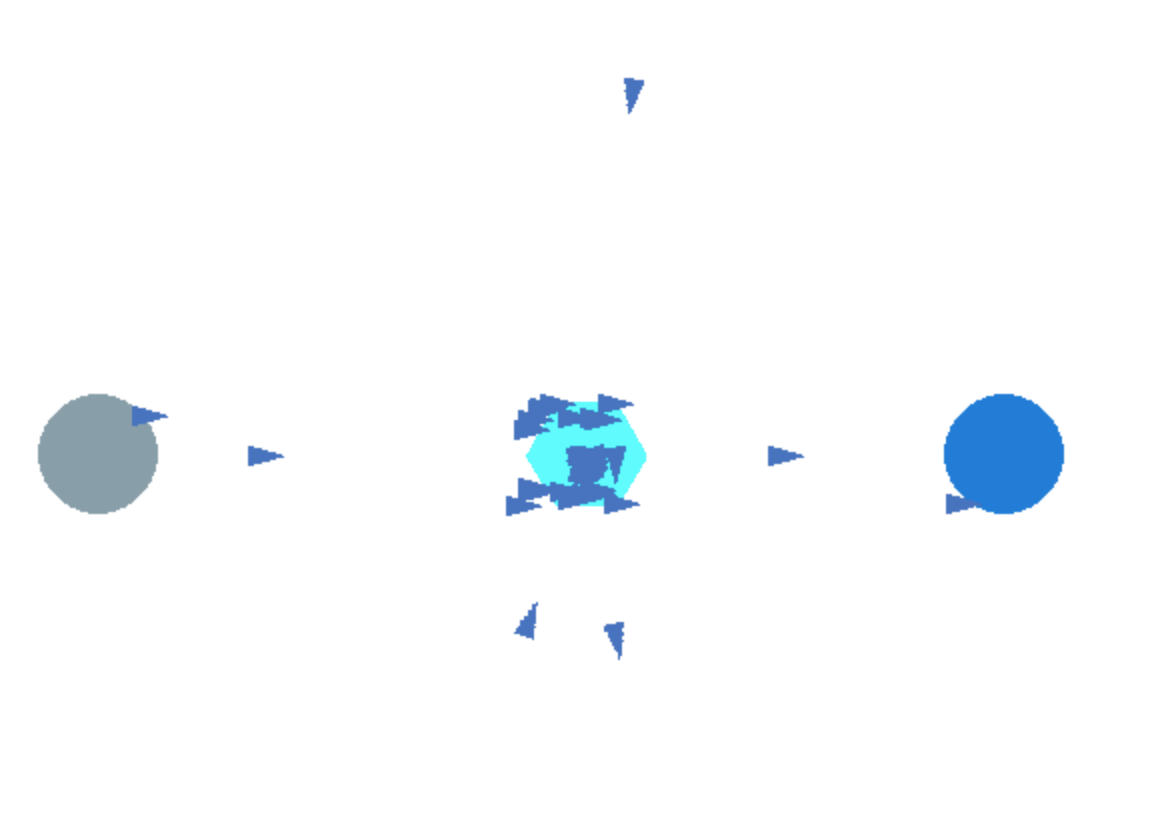}
         \caption{Best-of-N problem with two sites (circles) and 50 agents (triangles). The hexagon represents the hub.}
         \label{fig:exampleenv}
\end{figure}

The best-of-N problem is illustrated in Fig.~\ref{fig:exampleenv} for a problem with two sites, fifty agents, and a hexagonal hub. The agents, which are represented as triangles pointing in their direction of travel, explore the world. When they find a site of potential interest, they return to the hub and inform other agents. If they fail to find a site, they return to the hub and observe other agents. Agents travel between a site of interest and the hub to assess the site and to recruit other agents to the site. Agents recruiting for a site can sense when a quorum of agents are at the hub, and when a quorum is reached the collective decides that the site is the best solution to the problem.

\subsection{Agent Based Model (ABM)}

\begin{figure}[h]
      \centering      
      \includegraphics[width=0.95\linewidth]{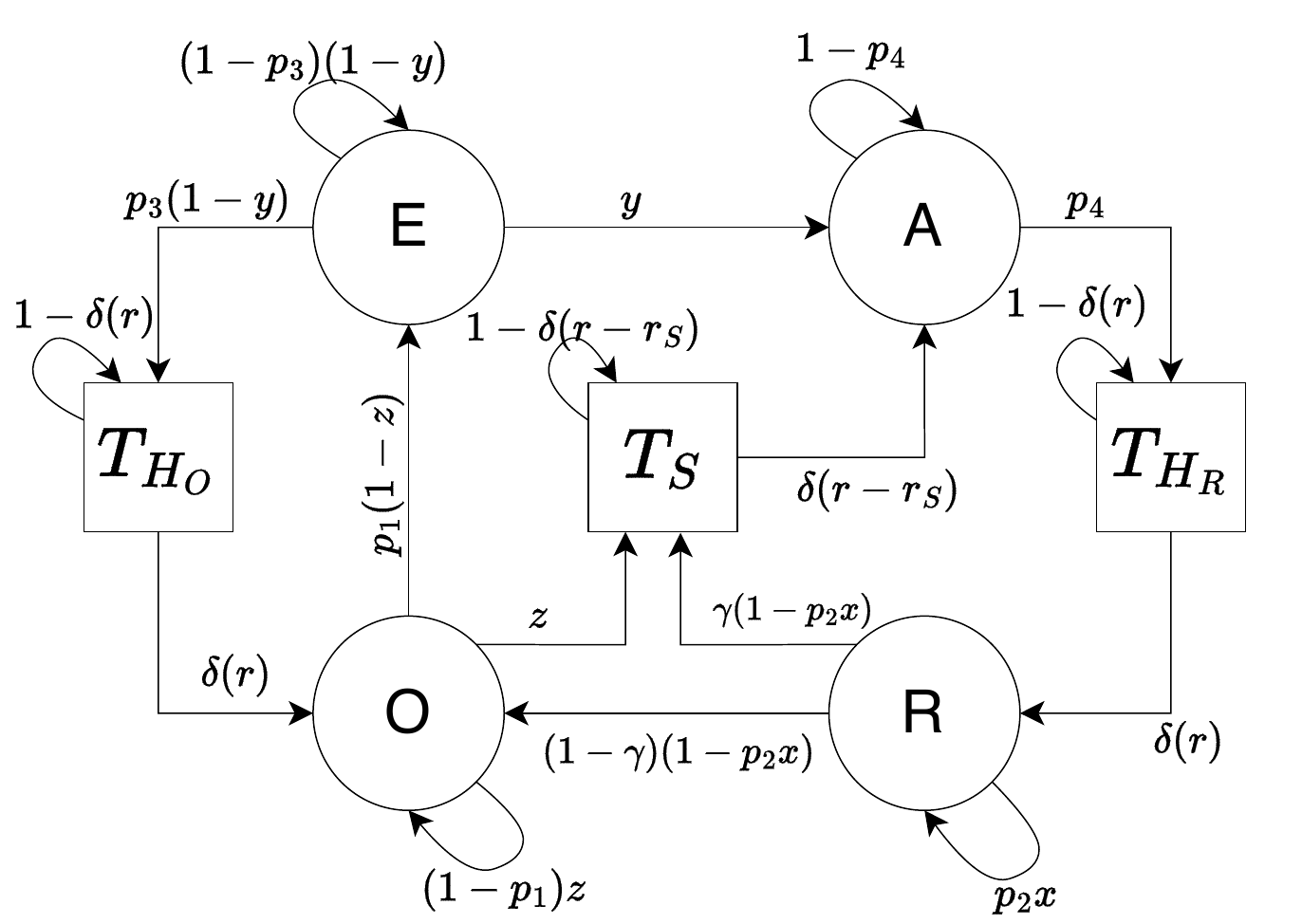}
         \caption{Agent-Based Model}
         \label{fig:main1}
\end{figure}

Unlike ABMs modeled as differential equations~\cite{cody_evaluation_2017,reina2015design} or finite state machines with augmented with extra memory~\cite{JainGoodrich2021a}, the ABM in this paper is  satisfies the Markov condition, where every agent's next state is dependent only on its current state. 
Each agent runs its copy of the state machine illustrated in Figure~\ref{fig:main1} with the following states. $O$: Observe, $E$: Explore, $A$: Assess, $R$: Recruit, $T_{H_O}$: Travel to Hub to Observe, $T_{H_R}$: Travel to Hub to Recruit, $T_S$: Travel to Site. $\delta$ is the dirac-delta function, $r$ is the position of the agent and $r_S$ is the position of a site. 

State transitions depend on the current state and position relative to the position of sites, hub and other agents.
The transition probabilities are shown on the edges. The transitions from the travel states ($T_{H_O}$, $T_{H_R}$, $T_S$) are represented by $\delta$ functions, which means that agents transition from these travel states only when they reach their destination, which is either the hub ($T_{H_O}$, $T_{H_R}$) or the site ($T_S$). 

The probabilities $p_1$, $p_2$, $p_3$, and $p_4$ are modeled as Bernoulli distributions with parameters set to control the mean time agents spend in $O$, $R$, $A$, $E$, and $A$, respectively. In the second set of experiments, samples from the Bernoulli distribution are obtained from numpy's \textsf{random binomial} function with number of trials as 1.
Transitions from the explore state depend not only on the Bernoulli distribution, $p_3$, but also on the probability that a site is discovered and it's quality ($y$). 
The transition from the recruit state ($R$) depends on (a)~the Bernoulli distribution, $p_2$, (b)~the number of times that an agent \textit{reassesses} the site, which is determined by quality of the site, $\gamma$, and (c)~the duration of recruiting, which is also a function of site quality, $x$.

The transition from the observe state depends on two parameters: whether the agent is recruited by another agent to assess a site, which is denoted by $z$, which is a function of recruiters for each site. The other parameter is how long the agent dwells in the observe state in the absence of being recruited, which is given by the parameter $p_1$. The site to which the agent is recruited is proportional to the number of agents recruiting to each site.





\subsection{Representing Collective State as a Tensor}
\label{sec:tensor}
A key insight from prior work is that information about the states of individual agents can be combined to create a compressed representation of collective state~\cite{JainGoodrich2021a,GoodrichJain2020b}. Unfortunately, that prior work was not sufficiently powerful to scale when sites could be at different lcoations in the world or when the number of agents changed. Consequently, this paper uses repesentation based on a relational database.

The relation header is the list of agent states ($R$, $D$, etc.) plus the quality of the site ($Q$) that the agent is traveling to, traveling from, recruiting to, or assessing. A unique agent identifier is also included, yielding a relation like Table~\ref{tab:relation}
\begin{table}[hbtp]
    \centering\begin{tabular}{|c|c|c|c|c|c|c|c|c|}\hline
        $Q$ & $R$ & $A$ & $T_{H_R}$ & $T_S$ & $O$ & 
 $E$ & $T_{H_O}$ &  $ID$ \\
 \hline\hline
1.0	& 0	& 0	& 0 & 1 & 0 & 0 & 0  & 2\\
0.5 & 1 & 0 & 0 & 0 & 0 & 0 & 0  & 0\\
0.5 & 0 & 1 & 0 & 0 & 0 & 0 & 0  & 3\\
0 & 0 & 0 & 0 & 0 & 0 & 0 & 1  & 1 \\
\hline
    \end{tabular}
    \caption{Relation representing the individual states.}
    \label{tab:relation}
\end{table}


Table~\ref{tab:relation} represents a collective with four agents. Agent~2 is traveling to a site with quality $q(s) = 1.0$ to assess it, agent~0 is recruiting to site with quality $q(s) = 0.5$, agent~3 is assessing a site with quality $q(s) = 0.5$, and agent~1 is returning home after failing to discover a site while exploring. In effect, each agent is represented by a one-hot encoding of the state the agent is in, augmented with the quality of a site an agents is favoring and agent identifier.


Tuples in the relation can be sorted by the values  of the tuples and then concatenated together into a tensor of tuples after removing agent~ID. This tensor is an anonymized representation of collective state. In a subsequent section, we explore what happens when we provide global information to the collective for large numbers of agents. This global information is appended to the start of the tensor.

\subsection{Representing State Dynamics as A Graph}
The collective state graph is constructed by creating a node for each tensor, and creating an edge between nodes if the swarm can evolve from the collective state in one node to the collective state in the other. Each tensor encodes the features associated with each node in the collective state graph. Graph edges encode transitions between collective states. Some nodes can transition to multiple next states, and the probability of the specific transitions is determined by the probabilities with which agents transitions between states in their individual state machines.

\begin{figure}[h]
      \centering
    \includegraphics[width=0.9\linewidth]{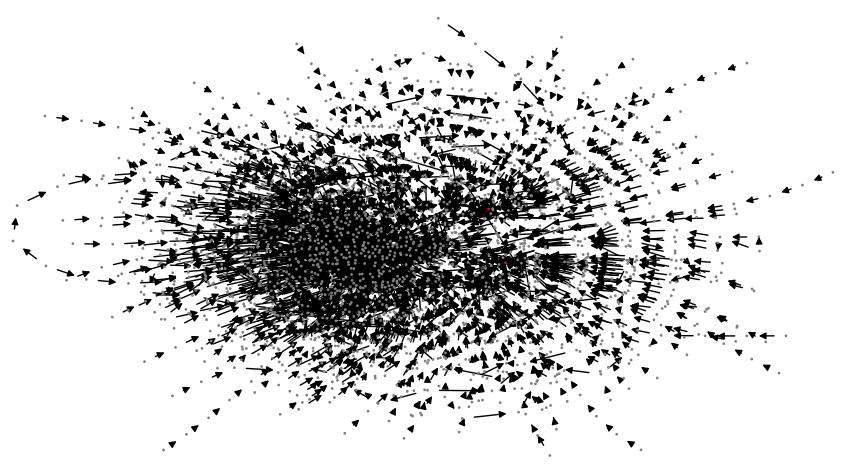}
         \caption{Example trajectories generated from 1500 trials for a problem with ten agents and two sites. Each point represents a unique tensor, and each arrow is the edge between tensors. The visualization is made using the graphviz visualization method in networkx. }
         \label{fig:Trajectories_simple_collective}
\end{figure}
Fig.~\ref{fig:Trajectories_simple_collective} represents a collective state graph for a network with 10 agents and two sites. Each point in the graph indicates a unique tensor, and each edge indicates a transition from one tensor to another. This graph was constructed by starting each agent at a random initial state, collecting 1500 trials, and keeping the largest (weakly) connected component.

\subsection{Information in the Tensor}
It is useful to explore what kind of information can be derived from the collective state tensors for a very small collective.  Consider a collective with only ten agents and two sites, one site with maximum quality $q(s_1)=1$ and the other site with relatively low quality, $q(s_2) = 0.5$. Because there are only a few agents and sites, the number of possible tensors is (relatively) small, so running several simulations provides a reasonable approximation of the entire graph.

\begin{figure}[htbp]
      \centering
    \includegraphics[width=0.95\linewidth]{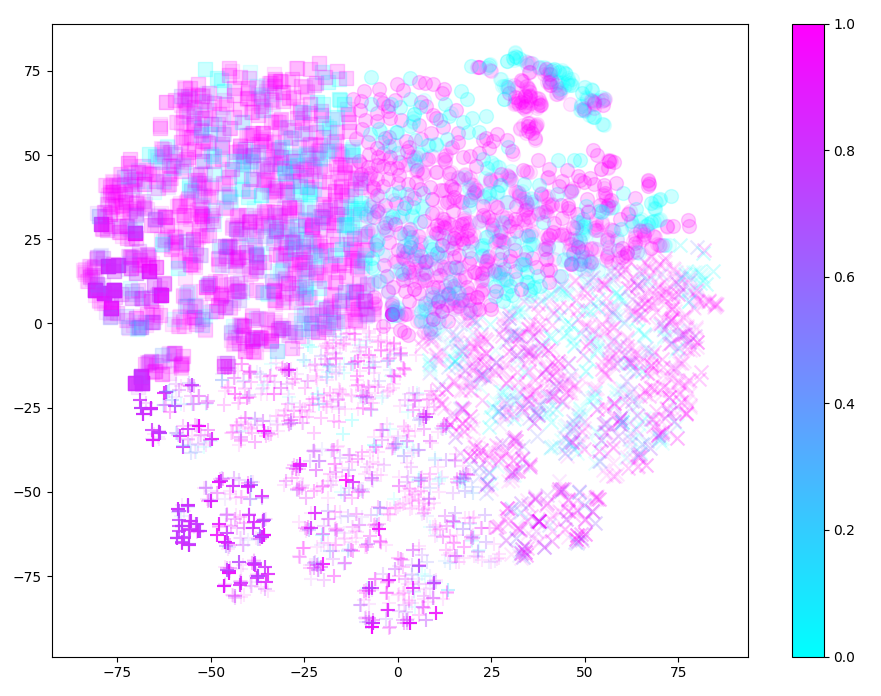}
         \caption{Clustering of 2D embedding using t-SNE~\cite{hinton2002stochastic}, and correlation of cluster with success probability. There were 10 agents, two sites ($q(s_1)=1$, $q(s_2)=0.5$) and a quorum thresold of 2 agents. Bernoulli parameters were set based on the mean times in a state: $O$ 8sec, $A$ 3sec, $R$ $6q(s)$sec. The number of reassessing trips was $\propto 3q(s)$. The Bernoulli parameter for being recruited by a single recruiting agent was~40sec. Explore agents used $y=\delta(D/2)$ so agents deterministically stopped exploring when they reached half the world dimension. Sites were placed at $D/4$ from the hub.  }
         \label{fig:simple_embedding}
\end{figure}

\begin{figure*}[!ht]
      \centering
    \includegraphics[width=0.95\linewidth]{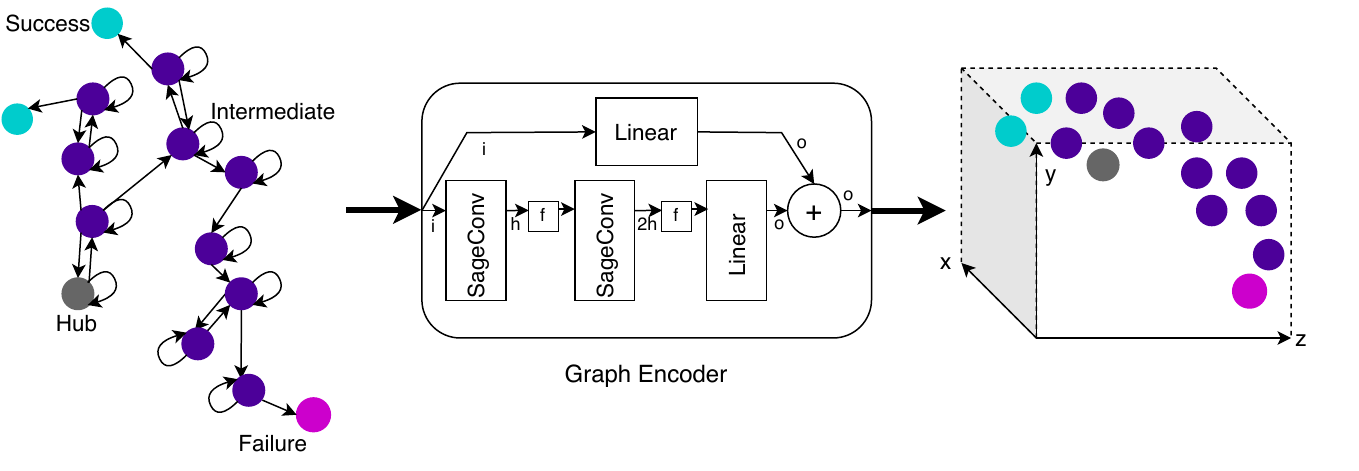}
         \caption{Encoder architecture: \textsf{i} is the input dimension, \textsf{h} is the hidden dimension, \textsf{o} is the output dimension, and \textsf{f = ReLU} is the activation function.}
         \label{fig:GNN}
\end{figure*}

We ran 500 trials with agents placed in random starting states and locations in the world that were appropriate for the state (e.g., moving in the world if in an explore state). The decision quorum threshold was set to three agents, which yielded 348 trials in which the agents chose the best site and 152 trials where they chose the inferior site. A tensor was part of a successful trajectory if (a)~the trajectory ended in choosing the best site and (b)~the tensor was visited at least once in the trajectory. The probability that a tensor yielded success was the number of times a tensor was part of a successful trajectory divided by the number of times it was part of any trajectory.

The shapes in Figure~\ref{fig:simple_embedding} represent different clusters of tensors. The clusters were computed (a)~by applying the t-SNE algorithm~\cite{hinton2002stochastic} to compress each tensor into a two dimensional embedding and then (b)~applying $k$-means clustering to form four clusters, indicated by shapes: $+$, $\bigcirc$, $\times$, and $\square$. Colors represent the probability that the tensor was part of a successful trajectory. Low alpha-values were used for tensors visited fewer than 10\% of the trajectories to indicate unreliable success estimates.  

Figure~\ref{fig:simple_embedding} can be interpreted as a success surface, represented by the heat map superimposed over the 2D embedding. The surface moves from high success in the lower left to low success in the top right. The $+$'s indicate tensors that were seen often and were usually part of successful trajectories. The $\square$'s and  $\times$'s indicate tensors that had fewer successes or were visited less often. The $\bigcirc$'s indicate tensors that rarely appeared on successful trajectories. 
Qualitatively, there is a positive correlation between the clusters and the success probability. This indicates that there is information in the tensors about the probability that a collective state will yield success. The next section uses this information to create low dimensional embeddings via graph encoding.



\section{Low-Dimensional Embeddings}
The previous section assumed the entire graph was known, which is unreasonable when there are many agents, sites, or possible site locations. This section addresses this limitation by using a GraphSage based graph encoder~\cite{hamilton2017inductive} to inductively learn the graph embedding.


\subsection{Input State Tensor for Embeddings}
Three modifications from Table~\ref{tab:relation} in Section~\ref{sec:tensor} are made to the relational database and tensor. First, rather than using a one-hot encoding for agent state, agent state, denoted by $S$,  is represented as a floating point value given by the following: $R=0/6$, $A=1/6$, $T_{H_R}=2/6$, $T_S = 3/6$, $O=4/6$, $E=5/6$, and $T_{H_O}=6/6$. Second,
unlike the previous experiment where the graph only applied to two sites that were at fixed locations, the graph encoder needs to work for sites at different locations. Thus, the $(x(s_i),y(s_i))$ position of the site $s_i$ favored by the agent, if any, is added to the agent record, after normalizing by maximum distance of the environment. Each record is therefore a tuple $[q(s_i), S, x(s_i), y(s_i)]$. Third, unlike the previous experiment where the number of agents was fixed, the graph encoder needs to work for different numbers of agents, which we bound to be less than or equal to 10. The tuple is constructed as described above, but the record for any ``extra" agents is the constant $[0, 1, 1, 1]$. 



\subsection{Graph Convolutional Neural Network}

We train the network inductively by forming subgraph samples. A subgraph sample is formed by creating the nodes and edges from a single single simulation. 
The encoder architecture is illustrated in Figure~\ref{fig:GNN}. The input dimension \textsf{i} is $40$, the hidden dimension \textsf{h} is $20$, and the output dimension \textsf{o} is 3, which yields a 3-dimensional embedding.

We leverage the GraphSAGE convolution layers~\cite{hamilton2017inductive} for aggregating features from a node's neighbors, thus enabling the learning of rich and complex node embeddings. The architecture consists of two graph convolution layers followed by an activation function ReLU, and a third Linear layer. We also incorporate a residual connection, first introduced in~\cite{he2016deep}, and used widely in LSTMs~\cite{hochreiter1997long}, transformer based systems like GPT-3~\cite{brown2020language} and AlphaFold~\cite{jumper2021highly}. This directly connects the input to the final output through a linear transformation to match the output dimensions. This shortcut is added to the output after the third convolution, facilitating an element-wise addition that merges the transformed input directly with the learned features. This residual mechanism is crucial for alleviating the vanishing gradient problem in deep neural networks, enabling the model to preserve information from the input throughout the network to enhance learning by providing alternate pathways for gradient flow.

\subsection{Loss Function}
The loss function used to train the neural network is based on graph autoencoders~\cite{ahn2021variational,pan2018adversarially} which aim to create graph embeddings in which nodes that are adjacent in the network have embeddings that are close together. This is done by taking the encoding vectors for two nodes, ${\mathbf x}$ and ${\mathbf y}$, and maximizing the sigmoid of the cosine similarity measure \begin{equation}
    \sigma({\mathbf x}^T{\mathbf y}).
    \label{eq:encoder_output}
\end{equation} When the embeddings of the two nodes are close to (far from) each other the output of Eq.~(\ref{eq:encoder_output}) is close to one (close to zero).

Thus, the output of Eq.~(\ref{eq:encoder_output}) approximates the existence of an edge in the original adjacency matrix. A binary cross entropy loss function with logits~\cite{bceloss} is used to compute the difference between the 0's and 1's in the adjacency matrix and the 0's and 1's approximated by Eq.~(\ref{eq:encoder_output}). In essence, this approach penalizes the model when it fails to align its perceived similarities with the provided adjacency criteria, guiding it to learn an embedding space where the desired relationships are accurately captured.

\section{Experiment Design}
This section addresses the following research questions.
\noindent\textbf{Problem 1}: \textit{Can useful 3D embeddings be generated for multiple different environment and agent configurations?} 
\noindent\textbf{Problem 2}: \textit{Do the embeddings for Success, Failure and Hub conditions exhibit useful clustering?}



\subsection{Experiment Conditions}
\label{sec:exp}

\begin{table}[h]
    \centering
    \begin{tabular}{|c|c|}
     \hline
\textbf{Parameter} &\textbf{Values} \\
\hline
\hline
x & $2/(2 + e^{-7q})$ \\
\hline
y & $q\delta(r-r_S)$ \\
\hline
z & $\delta(p_r)/|R|$ \\
\hline
$p_r$ & binomial($|R|$,0.1) \\
\hline
$p_1$ & binomial(1,0.01) \\
\hline
$p_2$ & binomial(1,0.99) \\
\hline
$p_3$ & binomial(1,0.02) \\
\hline
$p_4$ & binomial(1,0.1) \\
\hline
$\gamma$ & $q^{0.5}$ \\
\hline
Threshold $\tau$ & 0.5 \\
\hline
Constraints of  & $|q_{s_1}-q_{s_2}| < 0.5$ \\
Quality of Sites & min$(q_{s_1},q_{s_2}) > 0.5$ \\
\hline
Simulation run-times $T$ &  $T \in  \{1000, 10000, 35000\}$\\
\hline
Distance of Sites & $d_{site} \in \{100, 150, 200\}$\\
\hline
Maximum Distance & $1000$\\
\hline
Number of Agents & $K \in \{5, 10\}$\\
\hline
Number of Sites & $N \in \{2, 3, 4\}$\\

\hline
    \end{tabular}
    \caption{Parameters for the ABM and Simulations.}
    \label{tab:params}
\end{table}


Our environment and swarm configurations consist of defining agents, sites, qualities and distances. We present the parameters used in the ABM and the constraints for our simulations in the Table~\ref{tab:params}. The convergence criteria is set by a threshold of agents recruiting for a given site at the hub.

We start with three configurations: Condition 1 has 100\% Observe, Condition 2 has 50\% Explore and 50\% Observe, and Condition three has 90\% Observe 10\% Recruiting for worst site. Three trajectories are produced by running one simulation for each configuration. 10 tensors from these trajectories are randomly selected to serve as initial conditions.  
For each initial conditions from the previous paragraph we run 10 simulations for all possible pairs of distances, qualities, and runtimes from Table~\ref{tab:params}. The parameters in Table~\ref{tab:params} are chosen so that not all trajectories end in success.

The convergence criteria is set by the quorum threshold as $\tau*K$, where $K$ is the number of agents. Therefore, if more than $\tau$\% agents (6 for 10 agent colonies, and 3 for 5 agent colonies in our case) are recruiting for the same site at the Hub at the same time, the simulation ends. If the simulation is unable to reach convergence criteria, we do not consider its time to converge in the calculations.

\subsection{ABM Results}

\begin{figure}[h]
      \centering
    \includegraphics[width=0.95\linewidth]{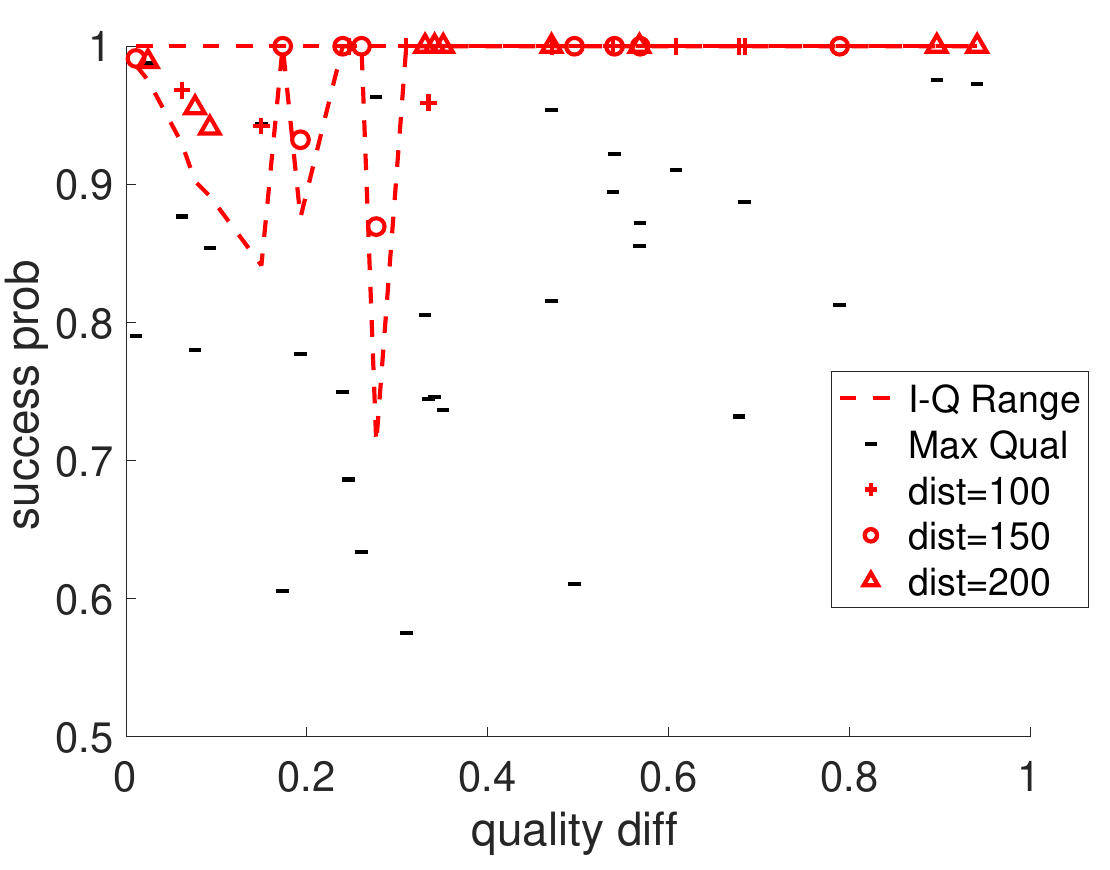}
         \caption{Mean Site quality difference vs Success with inter-quartile range}
         \label{fig:succABM}
\end{figure}

\begin{figure}[h]
      \centering
    \includegraphics[width=0.95\linewidth]{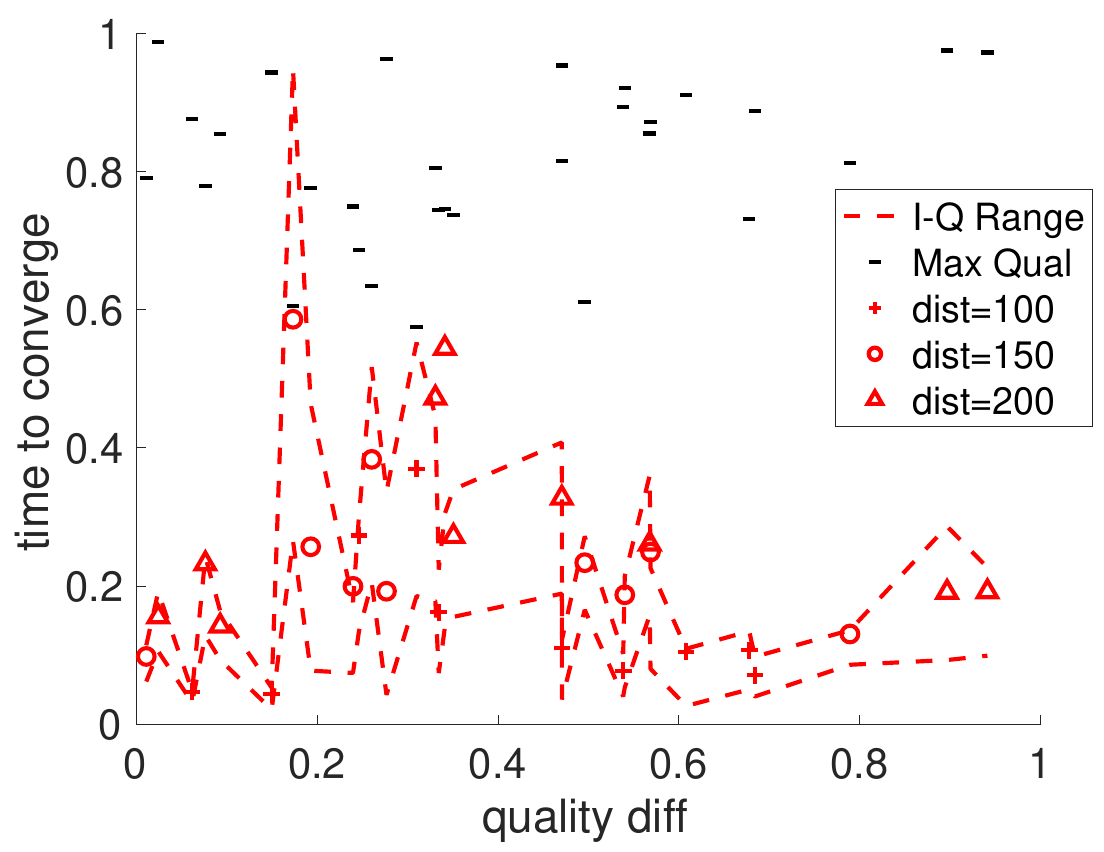}
         \caption{Mean Site quality difference vs Time with inter-quartile range}
         \label{fig:timeABM}
\end{figure}

Figure~\ref{fig:succABM} and~\ref{fig:timeABM} show the results of success and time to converge, respectively, for the simulations with 2 sites and 10 agents. The dashed line denotes the inter-quartile range, and the markers show the mean values. A $+$ denotes a distance of 100, $\circ$ is 150, and $\triangle$ is 200. The black $-$ denotes the maximum quality among the sites in that simulation.
The success metric if we choose site $i$ is defined as $q(s_i)/max(q_{s_1}, q_{s_2})$. 

Figure~\ref{fig:succABM} shows the success vs quality difference for our simulations, for different distances. We see that when the difference between site qualities is low (left side of the plots), the success is still high, since it doesn't matter which site you choose. The success is also high when one site has a much higher quality than the other (right side of the plots). In the middle, the success drops. Site distance ($+$, $\circ$, $\triangle$) have very little affect on the success metric.

Figure~\ref{fig:timeABM} shows that distance affects time to converge, as expected. The farther the sites, the higher the time to converge. Both the differences of site quality ($x$-axis) and the max quality (black dashes) affect the time to converge. When the quality difference is 0.2, decreasing site quality corresponds to increasing convergence time. By contrast, when quality difference is 0.5, convergence time is small for all values of maximum site quality. 




\subsection{Labeling Nodes}

From the simulations that converge, nodes where the better site among the two sites is chosen at the convergence are marked as ``Success" (cyan),  and those where the worse site is chosen are marked as ``Failure" (magenta). Nodes where no agent is \textit{site-oriented} are marked as ``Hub" (black). Every other node is marked as ``Intermediate" (purple).

\section{Experiment Results and Discussion}

This section presents the data from the experiments. 

\subsection{Results}
\begin{figure}[!h]
    \centering
    \includegraphics[width=\linewidth]{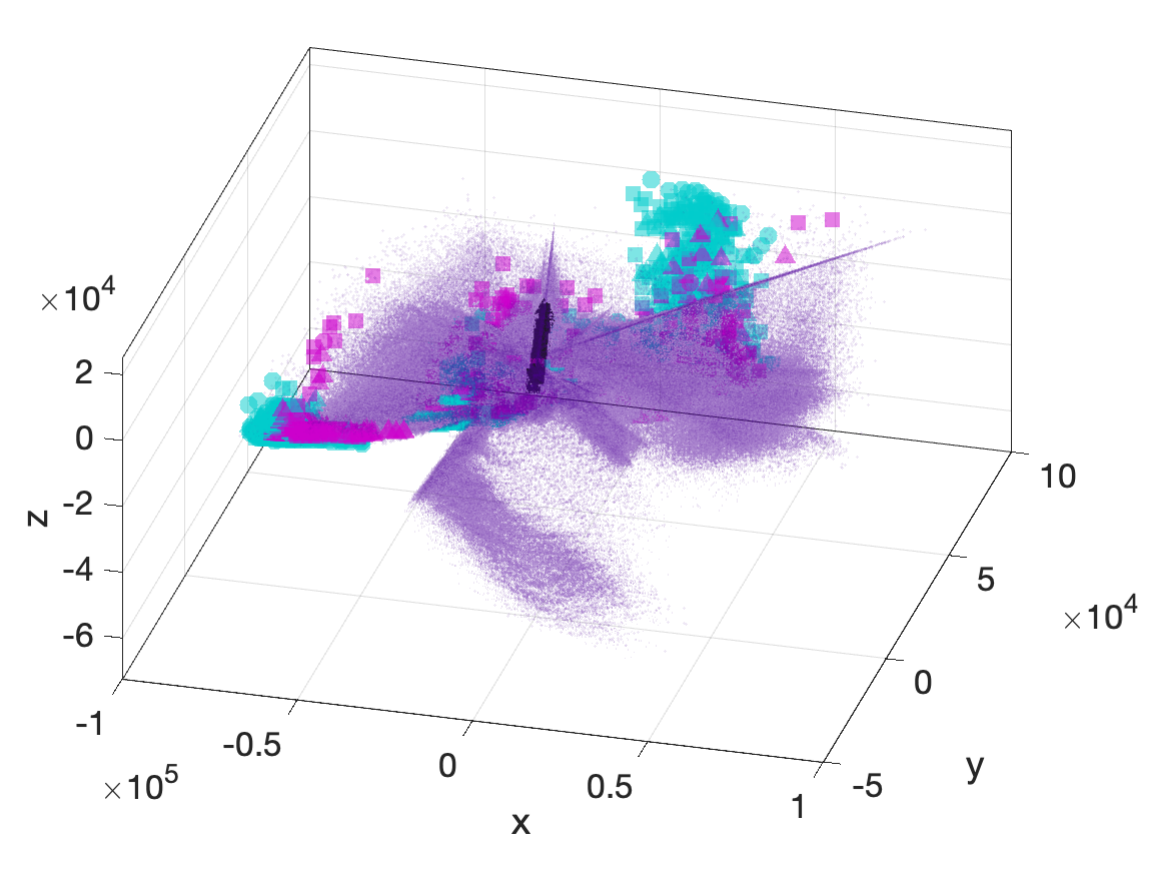}
    \caption{3D Embedding for varying environments and number of agents. }
    \label{fig:localembedding}
\end{figure}

Figure~\ref{fig:localembedding} shows the 3D embeddings for all tensors in the experiments. The smaller cyan and magenta markers denote simulations with 5~agents, and the larger markers denote 10 agents. The $\square$ markers, $\triangle$ markers, and $\bigcirc$ markers indicate conditions with two sites, three sites, and four sites, respectively. It is difficult to see in the figure, but there is very little difference between the embeddings for conditions that have 5 agents and conditions that have 10 agents.

The embeddings in  Figure~\ref{fig:localembedding} provide information about which tensors appear on multiple \textit{trajectories}. Transparency for the intermediate (non-hub, non-success, non-failure) embeddings in purple is set to highlight that frequently encountered trajectories tend to aggregate together; nodes encountered more frequently are darker. Frequently encountered embeddings tend to aggregate, and these aggregations correspond to probable trajectories of the collective. Infrequent trajectories  correspond to sparse point areas.

\subsection{Discussion of Research Questions}
The embeddings in Figure~\ref{fig:localembedding} also indicate that embeddings encountered on successful (failed) trajectories tend to cluster with other successful (failed) embeddings. Thus, the proximity of an embedding should be useful for predicting whether the corresponding tensor is likely to yield success or failure. Unlike Fig~\ref{fig:simple_embedding}, only successful or failed outcomes are shown, not the probability of success. Thus, areas where magenta and cyan markers overlap indicate uncertain outcomes.



The results suggest that useful lower dimensional (3D) embeddings can be generated for a system with varying number of agents and sites, which means that the answer to the first research question is, subjectively, yes. The embeddings show the likely paths taken by the colony to reach either success or failure, or not converge in some cases. 

Subjectively, we see the potential for finding clusters corresponding to basins of attractions: the ``hub" region, the ``success" region, the ``failure" region, and the ``intermediate" region. This suggest that the answer to the second research questions is, subjectively, yes. Importantly, there is not a single cluster for these areas of interest, but multiple clusters in the embedding space (multiple basins of attraction). Additionally, there are some regions where it is likely that the probability of success is low, which may require additional information from the world to disambiguate.

Given these observations about how frequently encountered embeddings aggregate and how embeddings for  successful and failure trajectories cluster, we  speculate that using representations like the one shown in Figure~\ref{fig:localembedding}, can used to predict swarm behavior. This information could potentially enable a human to help regulate the swarm behavior. 

\section{Future Work}


The results suggest that embeddings work for 5 and 10 agent groups, but future work should include more agents and experiments with more world configurations. Future work should also explore the ``harder" problem of differentiating between varying levels of the probability of success. Next, future work should take advantage of edge weights, which can serve as explicit transition probabilities and could lead to richer embeddings. Another direction to explore is various types of global and agent state information in the state tensor to solve different downstream tasks. 
With the growing developments in transformers, it would also be reasonable to look at the performance of transformers solve this problem.  


\addtolength{\textheight}{-3cm} 


\section{Acknowledgements}

The work was supported by the US Office of Naval Research under grant N00014-21-1-2190. The work does not represent opinions of the sponsor. We thank Chaitanya Dwivedi at Amazon AGI for his suggestions on the graph convolutional neural network architecture.

\bibliographystyle{ieeetr}
\bibliography{sample.bib}

\end{document}